# Magnetic field-induced quantum phase transitions in a van der Waals magnet

Siwen Li[1,+], Zhipeng Ye[2,+], Xiangpeng Luo[1,+], Gaihua Ye[2], Hyun Ho Kim[3], Bowen Yang[3], Shangjie Tian[4], Chenghe Li[4], Hechang Lei[4], Adam W. Tsen[3], Kai Sun[1], Rui He[2,*], and Liuyan Zhao[1,*]

[1]Department of Physics, University of Michigan, 450 Church Street, Ann Arbor, Michigan 48109, USA

[2]Department of Electrical and Computer Engineering, 910 Boston Avenue, Texas Tech University, Lubbock, Texas 79409, USA

[3]Institute for Quantum Computing, Department of Chemistry, and Department of Physics and Astronomy, University of Waterloo, Waterloo, 200 University Ave W, Ontario, N2L 3G1, Canada

[4]Department of Physics and Beijing Key Laboratory of Opto-electronic Functional Materials & Micro-nano Devices, Remin University of China, Beijing 100872, China

[+] authors contributed equally;   [*] corresponding to: rui.he@ttu.edu and lyzhao@umich.edu

Exploring new parameter regimes to realize and control novel phases of matter has been a main theme in modern condensed matter physics research. The recent discovery of two-dimensional (2D) magnetism in nearly freestanding monolayer atomic crystals[1,2] has already led to observations of a number of novel magnetic phenomena absent in bulk counterparts[3-13]. Such intricate interplays between magnetism and crystalline structures provide ample opportunities for exploring quantum phase transitions in this new 2D parameter regime. Here, using magnetic field and temperature dependent circularly polarized Raman spectroscopy of phonons and magnons, we map out the phase diagram of chromium triiodide ($CrI_3$) that has been known to be a layered antiferromagnet (AFM) in its 2D films and a ferromagnet (FM) in its three-dimensional (3D) bulk. We, however, reveal a novel mixed state of layered AFM and FM in 3D $CrI_3$ bulk crystals where the layered AFM survives in the surface layers and the FM appears in deeper bulk layers. We then show that the surface layered AFM transits into the FM at a critical magnetic field of 2 T, similar to what was found in the few layer case. Interestingly, concurrent with this magnetic phase transition, we discover a first-order structural phase transition that alters the crystallographic point group from $C_{3i}$ (rhombohedral) to $C_{2h}$ (monoclinic) and thus, from a symmetry perspective, this monoclinic structural phase belongs to the 3D nematic order universality class. Our result not only unveils the complex single magnon behavior in 3D $CrI_3$, but also settles down the puzzle of how $CrI_3$ transits from a bulk FM to a thin layered AFM semiconductor, despite recent efforts in understanding the origin of layered AFM in $CrI_3$ thin layers [10,11,13-16], and reveals the intimate relationship between the layered AFM-to-FM and the crystalline rhombohedral-to-monoclinic phase transitions. These findings further open up opportunities for future 2D magnet-based magneto-mechanical devices.

Chromium triiodide ($CrI_3$) stands out amongst the large family of van der Waals (vdW) ferromagnets (FMs) investigated so far, because its isolated atomic crystal is one of the first two to realize two-dimensional (2D) ferromagnetic long-range order[1,2] in the monolayer limit, and more importantly, is the first-ever interlayer antiferromagnetic (AFM) semiconductor[8,9,17,18] in its multilayer form. This also makes $CrI_3$ distinct from the 2D intralayer AFMs [19-21]. From an application perspective, such unique magnetic properties of $CrI_3$ have already drawn enormous interest in developing novel device functionalities that are tunable by either magnetic[8,9,17,18] or electric[7] fields. From a fundamental science point of view, extensive static and dynamic measurements in both three-dimensional (3D) bulk[22,23] and 2D film[8,9,17,18,24,25] $CrI_3$ have been performed to understand the mechanism of the magnetism that has remained elusive until now. Our study here approaches the nature of $CrI_3$ magnetism through exploring the magnetic field dependence of spin wave excitations in $CrI_3$ and interrogating the interplay between the magnetic order and the crystalline structure.

In its bulk form, $CrI_3$ goes through a monoclinic ($C2/m$) to rhombohedral ($R\bar{3}$) structural phase transition at $T_s$ = 220 K and develops a FM long range order at $T_{FM}$ = 61 K[22]. Across $T_s$, the major structural change involves the shearing of the vdW bonded $CrI_3$ layers of the honeycomb lattice from a tilted to an aligned ABC stacking sequence[22]. Below $T_{FM}$, all the spin moments within and between layers align along the out-of-plane direction[22] and a spin wave gap of ~ 1 meV emerges at the Brillouin zone center Γ point as a result of the Ising-type exchange anisotropy[23]. The magnetic and crystallographic degrees of freedom couple strongly in bulk $CrI_3$ as evidenced by the enhanced reduction of interlayer spacing below $T_{FM}$[22]. In its thin layer form, the $CrI_3$ magnetism is known to survive below $T_{AFM}$ = 45 K, however, in a unique layered AFM order in which spins align along the same out-of-plane direction within each layer and alternate to the opposite orientation between adjacent layers[8,9,17,18], while its crystal structure is much less studied experimentally. The origin for this crossover from the FM in bulk to the layered AFM in thin films, as well as the role of the crystallographic structure in this transition, remains as outstanding open questions in the field of 2D magnetism.

In order to probe both the magnetic and crystallographic degrees of freedom in $CrI_3$, we carried out polarized Raman spectroscopy to detect the symmetry-resolved collective excitations of spin precessions (*i.e.*, magnons)[26] and lattice vibrations (*i.e.*, phonons)[27], respectively. To further reveal the interplay between these two degrees of freedom, we performed both temperature and magnetic field dependent Raman measurements covering a temperature ($T$) range from room temperature down to 10 K and a magnetic field ($B$) range from 0 up to 7 T. Because the stray magnetic fields cause the Faraday rotation of linearly polarized light transmitted through the objective in close proximity to the magnet[28], we chose circularly polarized light to perform reliable selection rule measurements (see Supplementary Information section 1 for Raman selection rules in the circular polarization basis for $CrI_3$, and see Methods for details of the Raman measurements).

We start by showing Raman spectra in both parallel and crossed circularly polarized channels, labeled as LL and LR in Fig. 1a, respectively, taken on a freshly cleaved 3D $CrI_3$ crystal at $T = 10$ K and $B = 0$ T, where LL(R) stands for incident and scattered light being left and left (right) circularly polarized, respectively. We categorize all the observable Raman modes into three categories based on their symmetry properties. The first category contains phonon modes of either $A_g$ or $E_g$ symmetry of the $C_{3i}$ point group as reported before using linearly polarized light[24,29,30]. Under the circular polarization basis in Fig. 1a, the $A_g$ phonon modes (labeled $P(A_g)$) only appear in the LL channel and the $E_g$ ones ($P(E_g)$) are solely in the LR channel, as expected for the rhombohedral crystal structure. The second group consists of two modes, $M_1$ and $M_2$, that were previously shown to be antisymmetric and attributed to surface magnetic excitations[24], showing up in the LL channel here. The third kind is exclusively very low frequency modes, $M_0$, that, however, have neither been detected experimentally[24,29,30] nor been predicted theoretically[31,32] before in Raman studies and are present in both LL and LR channels. Symmetry-wise, these $M_0$-type modes violate the selection rules for the rhombohedral crystal structure and the FM order. Energy-wise, their frequencies of ~ 4 cm$^{-1}$ (0.49 meV) are close to the reported bulk spin wave gap, which is on the order of 1 meV[23]. We note that the $M_0$ intensity is stronger in the anti-Stokes than the Stokes channel, possibly due to the resonance excitation to the charge transfer transition in $CrI_3$ and the broken time reversal symmetry of $M_0$.

To explore the nature of these potentially magnetism-related Raman modes, $M_{0-2}$, and investigate the magneto-elastic coupling between magnetic and phononic modes, we carried out careful out-of-plane magnetic field ($B \perp ab$) dependent Raman measurements at 10 K and show the results in the LL channel in Fig. 1b (see Supplementary Information section 2 for extended data of magnetic field dependent Raman spectra on 3D $CrI_3$ bulk). First of all, it is clear that the frequencies of $M_0$ scale linearly with the magnetic field, consistent with what would be expected for spin wave excitations, despite a discontinuity at ~ 2 T and a much more complex pattern for the magnetic fields lower than 2 T. In addition, at 0 T, $M_0$-type modes soften towards 0 cm$^{-1}$ when approaching the critical temperature $T_c$ of 45 K from below, which follows the trend of $\sqrt{T_c - T}$ as expected for the order parameter across the magnetic phase transition[33,34] (Fig. 1d top panel). Based on both the linear magnetic field dependence and the softening of $M_0$ frequencies, we confidently assign $M_0$-type modes to be spin wave excitations. Second, the $M_1$ and $M_2$ show no magnetic field dependence at all below 2 T and disappear at higher magnetic fields. Although their antisymmetric selection rule is indicative of their magnetic origin[26] and their temperature dependent intensities mimics closely that of the magnetic order parameter[33,34] (Fig. 1d middle and bottom panels), such a magnetic field independence of $M_1$ and $M_2$ before 2 T immediately rules out the possibility of them being conventional spin wave excitations. More insights about their potential nature will be discussed later on in Fig. 3. Third, the phonons exhibit contrasting selection rules below and above 2 T while their frequencies remain more or less independent of magnetic field. For example, the $E_g$ mode (at ~ 109 cm$^{-1}$) leaks into the LL channel (Fig. 1b) and the $A_g$ modes (at ~ 79 and 129 cm$^{-1}$) show up in the LR channel above 2 T (see Supplementary Information section 2).

Having described the Raman spectra evolution with increasing magnetic field, we show in Fig. 1c both Raman spectra in the LL and LR channels at our highest available magnetic field of 7 T. Comparing to the spectra taken at 0 T in Fig. 1a, one of the $M_0$ modes shifts up to nearly 9 cm$^{-1}$ with its spectral intensity primarily in the LR channel while neither $M_1$ nor $M_2$ are present in either the LR or LL channel. In addition, observable fractions of phonon intensities leak into the corresponding orthogonal channels, suggesting that the crystalline symmetry is lowered from the $C_{3i}$ point group at 0 T. So far, we have established that both the magnetic order and the crystalline structure change across a critical magnetic field $B_c$ of about 2 T. To gain more insights into these magnetic field-induced phase transitions and their relationships to one the other, in the following, we first discuss the magnetic phase transition from the magnetic field dependence of $M_0$ and then address the structural phase transition from the combination of the magnetic field dependence of the $M_{1-2}$, $A_g$ and $E_g$ phonons.

To provide a comprehensive picture of the magnetic field dependence of $M_0$, we summarize in Fig. 2a the key experimental result of the $M_0$ frequencies shifting as a function of the external magnetic field and the corresponding field-dependent spin wave calculations in the left and right panels, respectively. Here, we show the average $M_0$-type mode frequencies from the Lorentzian fit of the Raman spectra for both Stokes and anti-Stokes shift in both the LL and LR channels. Strikingly, despite the fact that bulk CrI$_3$ is considered as a simple Ising ferromagnet[22], we observe that three spin wave branches at magnetic fields lower than 2 T collapse into one across $B_c = 2$ T. In particular, a pair of the three branches below 2 T start with close frequencies of ~ 3.4 and 3.9 cm$^{-1}$ at 0 T and evolve in opposite trends at increasing magnetic field ($M_{0a}$ and $M_{0b}$), while the third one increases linearly since its appearance at ~ 1 T and continues after $B_c$ of 2 T with a weak discontinuity of frequency redshift ($M_{0c}$). The field dependence of $M_{0a}$ and $M_{0b}$ are typical behavior of spin waves in AFMs in which the angular momenta of two degenerate spin waves align parallel and antiparallel, respectively, to the external magnetic field, and $M_{0c}$ is consistent with spin waves in FMs.

Our experimental observations above suggest a mixed state of layered AFM and FM for a 3D CrI$_3$ bulk, in contrast to the literature assignment of a pure FM phase[22]. This is consistent with the absence of $M_{0c}$ below $B_c$ in CrI$_3$ flakes (see Supplementary Information section 3). As sketched in Fig. 2b, below $B_c$, top layers of bulk CrI$_3$ show layered AFM (denoted as sAFM) that is similar to what has been reported in 2D

CrI$_3$ thin films[8,9,17,18], and deeper bulk exhibits a FM order (denoted as bFM) that is consistent with what bulk magnetization measurements find. By performing numerical calculations of magnetic energy per Cr$^{3+}$ through long-range magnetic dipole-dipole interactions that favors interlayer FM, we find this energy scale is about 1 $\mu$eV[35], much weaker than the interlayer AFM exchange coupling of 150 $\mu$eV (see Supplementary Information section 4). Therefore, we rule out the possibility of pure magnetic energetic reasons and require surface reconstructions to establish the interlayer AFM coupling at surface while having FM coupling in bulk. Hence, the sAFM provides the pair of spin waves with opposite Zeeman shifts (M$_{0a}$ and M$_{0b}$) while the bFM leads to the third branch with a linearly increasing frequency with growing magnetic field (M$_{0c}$). Also noted is that the bFM provides an effective magnetic field of 0.32 T to lift the degeneracy of two spin waves of the sAFM at the 0 T external magnetic field. Above $B_c$, sAFM transits into FM in the same manner as layered AFM in CrI$_3$ thin flakes does[8,9,17,18], leading the entire CrI$_3$ crystal into a FM state as shown in Fig. 2b and resulting in a sole spin wave branch above $B_c$. This sAFM to FM transition in 3D CrI$_3$ bulk is further supported by a weak anomaly at $B_c = \pm 2$ T in the magnetization v.s. magnetic field measurements (see Supplementary Information section 5). Based on this proposed model, the calculated magnetic field dependence of the M$_0$-type mode frequencies (right, Fig. 2a) with a nearest-neighbor Ising spin Hamiltonian coincides well with our experimental data (left, Fig. 2a) (see the calculation details in Supplementary Information section 6). Based on the selection rule for M$_0$ evolving from the nearly equal weighted presence in both LL and LR channels at 0 T to a dominant selection in the LR channel at 7 T, we can further infer that the spin moments in CrI$_3$ orient from being tilted away from to nearly aligned with the surface normal as the external magnetic field increases.

Concurrent with the magnetic phase transition at $B_c$, we also observe an evident first-order structural phase transition manifested by abrupt changes in phonons (Fig. 3a-d). First and foremost, the appearance of fully symmetric A$_g$ phonons in the LR channel (Fig. 3a) reveals the transformation of their Raman tensors from the $\begin{pmatrix} a & . \\ . & a \end{pmatrix}$ to the $\begin{pmatrix} a & . \\ . & b \end{pmatrix}$ form, suggesting the loss of the three-fold rotational symmetry and thus the breaking of the rhombohedral crystal symmetry. We propose the shearing of vdW layers away from the aligned ABC stacking order (Fig. 3e), which is indeed a structural instability for CrI$_3$ bulk[22,36], as one means to transit from the C$_{3i}$ rhombohedral (nearly D$_{3d}$ because of weak interlayer interactions) to the C$_{2h}$ monoclinic crystal symmetry (see Supplementary Information section 1 for C$_{2h}$ Raman selection rules and Supplementary Information section 7 for its comparison to the high temperature monoclinic structure). Interestingly, this magnetic field-induced monoclinic phase mimics a 3D nematic long-range order with a director as its order parameter (indicated as elongated ellipses in Fig. 3e). It is known that such a 3D nematic order must emerge through the first-order phase transition[37] as is indeed our case here. Second, for the E$_g$ phonons, which are only present in the LR channel below $B_c$, some of their intensities exhibit leakage into the LL channel right across $B_c$ while some others show clear jumps in the LR channel at $B_c$ without leaking into the LL channel (examples of each case are shown in Fig. 3b and c). These two cases are consistent with the $E_g(C_{3i})$ phonons transforming into the $A_g(C_{2h})$ and $B_g(C_{2h})$ phonons, respectively, which further corroborates the proposed structural phase transition at $B_c$. Third, the antisymmetric $M_{1,2}$ modes stay nearly constant until $B_c$ and disappear right above $B_c$ (see $M_2$ in Fig. 3d and $M_1$ in Supplementary Information section 8). Both modes were initially interpreted as surface magnetic excitations based on their broken time reversal symmetry and thickness independence[24]. Here, we can immediately rule out the possibility that they are conventional bulk spin waves based on the magnetic field independence of their frequencies below $B_c$, and can confidently associate them with the sAFM because of their disappearance above $B_c$. Based on this, we propose one possible origin of $M_{1,2}$ to be a collective excitation made of two parts, one being the $c$-axis zone boundary phonon of the non-magnetic lattice $A(\vec{k}_c, \omega)$, and the other being the layered AFM order $M(-\vec{k}_c, 0)$, where $\vec{k}_c$ is the out-of-plane wavevector for the constituent phonon ($A$) and magnetism ($M$) at frequencies of $\omega$ and 0 (*i.e.*, elastic scattering off the layered AFM order), respectively. Such a collective excitation breaks the time reversal symmetry, possesses a total momentum of zero and becomes inaccessible in the FM phase above $B_c$ due to the finite momenta (see detailed analysis in Supplementary Information section 9). In other words, $M_{1,2}$ corresponds to a special zone-folding by a single copy (or more precisely, odd number of copies) of the magnetic order that breaks the time reversal symmetry. However, this proposed origin cannot account for the thickness independence reported in Ref. [24]. Further experiments are needed to pin down the exact nature of both $M_1$ and $M_2$ modes. Nevertheless, they are good indicators for the sAFM state and the rhombohedral crystal lattice (Fig. 3d and Fig. 1d). In contrast to conventional structural transitions, the emergence of this 3D nematic order is driven by an external magnetic field that has a much stronger coupling to electrons than to ions, supporting an electronic origin for this structural transition. One natural mechanism for it could be that this interlayer shear deformation increases the distance between the nearest interlayer spins and thereby reduces the exchange energy penalty for the field-induced layered AFM to FM transition. Magnetic field dependent Raman data on $CrI_3$ flakes show consistent results (see Supplementary Information section 10).

We then proceed to construct the temperature versus magnetic field phase diagram for bulk $CrI_3$ by performing temperature (magnetic field) dependent Raman measurements at multiple magnetic fields (temperatures). We show in Fig. 4a the temperature dependence of the $M_{0c}$ FM frequency at a series of magnetic fields, and we clearly observe a crossover between FM and paramagnetic (PM) states at ~ 65 K at all applied magnetic fields[37,38]. This crossover is represented by the striped line with the experimental data points from this work in Fig. 4c, and its extrapolation to zero magnetic field indeed corroborates the FM transition temperature of 61 K obtained from the magnetization measurements in 3D $CrI_3$ bulk[22].

Meanwhile, we display in Fig. 4b the magnetic field dependence of the $M_2$ intensity at several temperatures, and we discover a decreasing trend of the critical magnetic field $B_c$ with increasing temperatures. This shows that the novel mixed state of sAFM and bFM in bulk $CrI_3$ is bounded by a line of phase transitions of both the layered AFM to FM and the rhombohedral to monoclinic structures, which is highlighted by a solid brown line in Fig. 4c. Furthermore, based on the fact that the crystal structure of 3D $CrI_3$ bulk transits from the rhombohedral to monoclinic symmetry across ~ 220 K from below at 0 T[22], we propose the presence of a phase boundary for the paramagnetic rhombohedral phase so as to connect the two regions of monoclinic structures above 220 K at $B = 0$ T and below 45 K at $B > B_c$. Please note that while the starting point of this phase boundary is determined to be at $T = $ ~ 220 K, $B = 0$ T, its end point is unknown due to a lack of experimental data, and thus the gray dashed line in Fig. 4c only presents one possibility.

Our findings in 3D $CrI_3$ bulk of a novel mixed state of the surface layered AFM and the bulk FM and a magnetic field-induced first-order structural phase transition reveal a rich phase diagram for this vdW magnetic semiconductor, which unambiguously resolves the puzzling evolution from the FM in 3D to the layered AFM in 2D $CrI_3$. Furthermore, controlling this unique magnetism in vdW magnets and its interplay with crystalline structures opens up new possibilities for the realization of novel 2D magnetic phases and the applications in modern spintronics. While a broader class of vdW magnets with such unique magnetism and strong magneto-elastic coupling are to be identified, $CrI_3$ serves as an ideal platform to explore a variety of external controls, such as electric field, strain, and charge carrier doping, on the magnetism and its interplay with other degrees of freedoms.

**Methods**

**Growth of $CrI_3$ single crystals**     Single crystals of $CrI_3$ were grown by the chemical vapor transport method. Chromium power (99.99% purity) and iodine flakes (99.999%) in a 1:3 molar ratio were put into a silicon tube with a length of 200 nm and an inner diameter of 14 mm. The tube was pumped down to 0.01 Pa and sealed under vacuum, and then placed in a two-zone horizontal tube furnace. The two growth zones were raised up slowly to 903 K and 823 K for 2 days, and were then held there for another 7 days. Shiny, black, plate-like crystals with lateral dimensions of up to several millimeters can be obtained from the growth. In order to avoid degradation, the $CrI_3$ crystals were stored in a nitrogen-filled glovebox and exfoliated in dark to expose fresh surfaces right before the experiment (the air exposure time in a dark environment was less than 10s before the sample was sealed in the cryostat vacuum chamber).

**Micro-Raman spectroscopy**    Micro-Raman spectroscopy measurements were carried out using a 632.81 nm excitation laser with a full width hall maximum (FWHM) of 0.85 cm$^{-1}$, on resonance with the charge transfer and $Cr^{3+}$ $^4A_2$ to $^4A_1$ transitions of $CrI_3$ in order to increase the Raman sensitivity. According to the absorption measurements taken in $CrI_3$ flakes, the optical penetration depth in $CrI_3$ for 633nm is on the order of 30 layers. The laser beam on the sample site was focused down to ~ 3 μm in diameter and the laser power was kept at 80 μW, which corresponds to a similar fluence used in literature (10 μW over a 1-μm diameter area), to minimize the local heating effect. Backscattering geometry was used. The scattered light was dispersed by a Horiba LabRAM HR Evolution Raman microscope (1800 groves/mm grating) from Horiba Scientific, and detected by a thermoelectric cooled CCD camera provided by Horiba Scientific as well. Because of the small sample size on the order of a few microns, we used the confocal microscope geometry at a magnification of 40× to position the laser spot onto the sample of interest and perform Raman spectroscopy experiment. In order to suppress the reflected light and Rayleigh scattering, we added a notch filter with a central wavelength of 632.85 nm, with its transmission profile described in Table 1 below. A commercial variable temperature (< 10 K – 325 K), closed cycle, microscopy cryostat from Cryo Industries of America, Inc was interfaced with the Raman microscope. A commercial cryogen free room-temperature-bore (2'' in diameter and 6.88'' long) superconducting magnet from Cryo Industries of America, Inc was used to achieve the variable out-of-plane magnetic field from 0 T to 7 T. The cryostat cold finger, on which the samples were mounted, was inserted into the center of the room-temperature-bore of the magnet. In this work, the selection rule measurements were performed mainly under the circularly polarized basis to eliminate any artifacts of the Faraday effect from the microscope objective subject to the strong stray magnetic field. All thermal cycles were performed at a base pressure lower than $7 \times 10^{-7}$ torr.

| Transmission Efficiency | Stokes Edge (cm$^{-1}$) | Anti-Stokes Edge (cm$^{-1}$) |
|---|---|---|
| 10% | 3.7 | − 1.8 |
| 50% | 4.2 | − 2.2 |
| 90% | 7.8 | − 6.3 |

**Table 1 Transmission efficiency and their corresponding Stokes and anti-Stokes edges for the notch filter used in the experiment with a 632.81nm excitation laser.** The transmission efficiency is less than 10% in [-1.8, 3.7] cm$^{-1}$, between 10% and 50% over the ranges of [3.7, 4.2] and [− 2.2, − 1.8] cm$^{-1}$, and between 50% and 90% over the ranges of [4.2, 7.8] and [− 6.3, − 2.2] cm$^{-1}$.

**Acknowledgements**

We acknowledge helpful discussions with Xiaodong Xu, Roberto Merlin, Adam W. Tsen, and Elizabeth Drueke. L. Z. acknowledges support by NSF CAREER Grant No. DMR-1749774. R. H. acknowledges support by NSF CAREER Grant No. DMR-1760668 and NSF MRI Grant No. DMR-1337207. K. S. acknowledges support through NSF Grant No. NSF-EFMA-1741618. A. W. Tsen acknowledges support by NSERC Discovery grant RGPIN-2017-03815 and the Korea-Canada Cooperation Program through the National Research Foundation of Korea (NRF) funded by the Ministry of Science, ICT and Future Planning (NRF-2017K1A3A1A12073407). This research was undertaken, thanks in part to funding from the Canada First Research Excellence Fund. H. L. acknowledges support by the National Key R&D Program of China (Grant No. 2016YFA0300504), the National Natural Science Foundation of China (No. 11574394, 11774423, and 11822412), the Fundamental Research Funds for the Central Universities, and the Research Funds of Renmin University of China (15XNLQ07, 18XNLG14, and 19XNLG17).

**Author contributions**

L. Z. and R. H. conceived and initiated this project; S. T., C. L. and H. L. synthesized and characterized the bulk $CrI_3$ single crystals; Z. Y., G. Y., and R. H. performed the Raman measurements; S. L. performed the magnetic field dependent spin wave dispersion calculations under the guidance of K. S. and L. Z.; H. H. K., B. Y., and A. W. T. made the $CrI_3$ flake samples and performed magneto-tunneling resistivity measurements. S. L., X. L., and L. Z. analyzed the data; S. L., X. L., R. H. and L. Z. wrote the manuscript and all authors participated in the discussions of the results.

**Competing financial interests**

The authors declare no competing financial interests.

**Data availability**

The data that support the plots of this paper are available from the corresponding authors (R. H. and L. Z.) upon reasonable request.

**Figures and figure captions**

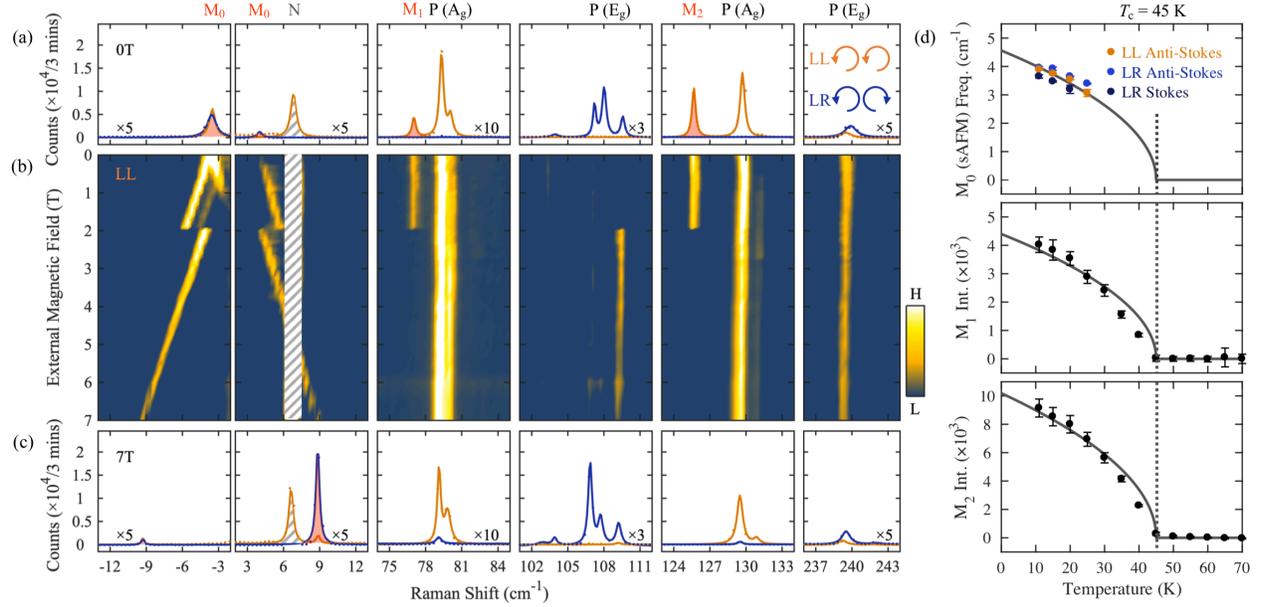

**Figure 1** Identifying the magnetism related Raman modes in 3D $CrI_3$. **(a)** Raman spectra taken at 10 K and 0 T in the LL and LR channels. LL(R) stands for the polarization channel in which the incident and scattered light is left and left (right) circularly polarized, respectively. Solid dots are raw data points and solid lines are Lorentzian fits. The multiplication factors to the spectra intensities are labeled in the corresponding panels. The Raman modes are labeled as $M_{0, 1, 2}$ for the magnetism related ones (shaded in orange), $P(A_g)$ and $P(E_g)$ for the phonon modes of $A_g$ and $E_g$ symmetries of the $C_{3i}$ point group, respectively, and N for the noise line from the incident light (shaded in gray stripes). Inset shows the legends for the polarization channels. **(b)** A color map of magnetic field dependent Raman spectra taken over a magnetic field range of 0 – 7 T at 10 K in the LL channel. The gray stripe-patterned shade is to block the noise line. **(c)** Raman spectra taken at 10 K and 7 T in the LL and LR channels. **(d)** Temperature dependence of $M_0$ frequency (including data for the LL anti-Stokes, LR anti-Stokes and Stokes shifts) and $M_{1, 2}$ intensities at zero magnetic field ($B = 0$ T) showing a clear onset at $T_c$ of 45 K. Error bars are defined as one standard error of the fitting parameters in the Lorentzian fit to the Raman modes.

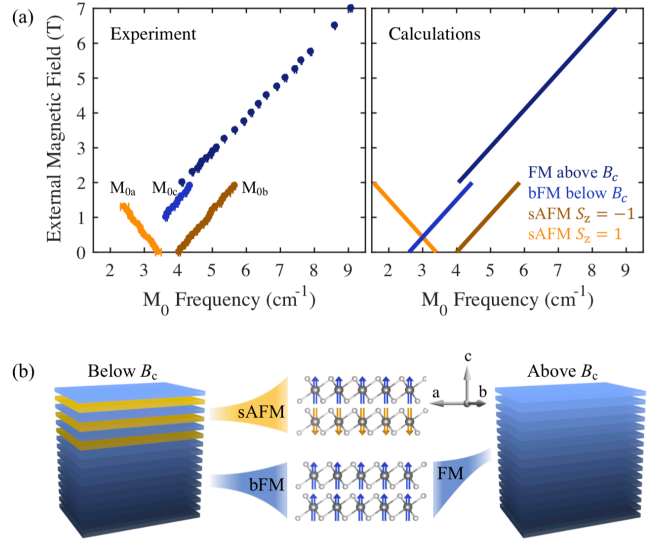

**Figure 2** **Establishing the mixed sAFM and bFM phase in 3D $CrI_3$.** **(a)** Experimental data (left) and spin wave calculations (right) for the magnetic field dependence of $M_0$-type mode frequencies. $M_{0a}$, $M_{0b}$, and $M_{0c}$ label the three spin wave branches below the critical magnetic field $B_c$ of 2 T. Error bars correspond to one standard error of the fitted central frequencies for the $M_0$-type Raman modes, and are mostly smaller than the data point symbols. **(b)** Schematic illustration for the mixed state of the surface layered AFM (sAFM for the state with alternating spin moments in the adjacent layers) and the deep bulk FM (bFM for the state with all the spin moments along one direction) below $B_c$ and the pure FM state above $B_c$.

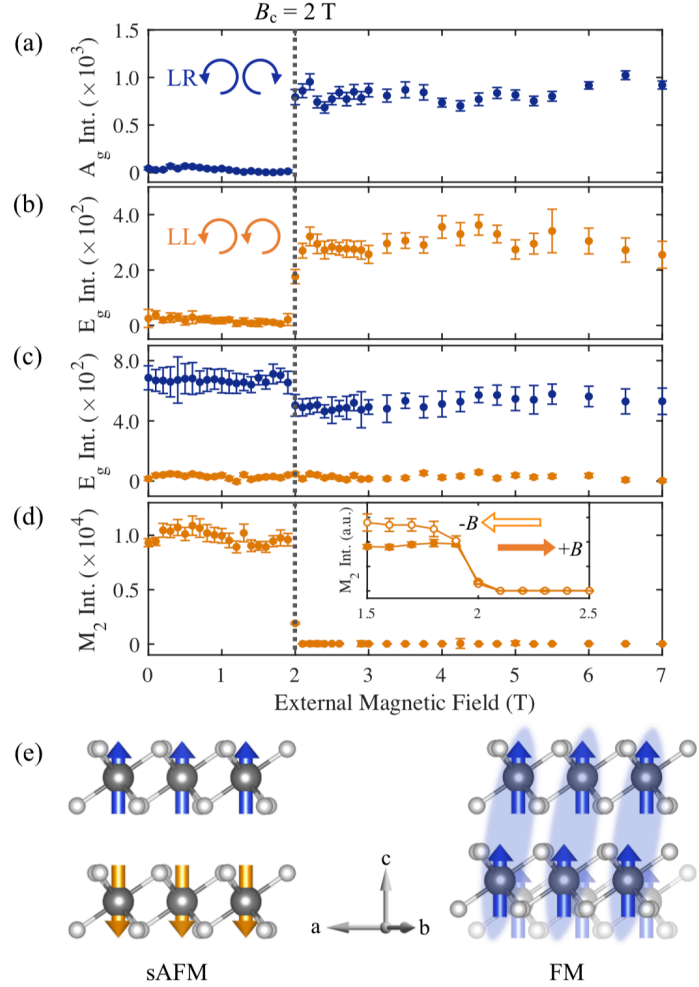

**Figure 3       Revealing the rhombohedral to monoclinic structural phase transition at $B_c$ in 3D CrI$_3$.** Magnetic field dependence of **(a)** a represent A$_g$ phonon (at ~ 129 cm$^{-1}$) intensity leakage into the LR channel, **(b)** an example of E$_g$ phonon (at ~ 109 cm$^{-1}$) intensity showing up in the LL channel, which corresponds to the E$_g$ phonon of the C$_{3i}$ structure (*i.e.*, E$_g$(C$_{3i}$)) transforming into the A$_g$ phonon of the C$_{2h}$ structure (*i.e.*, A$_g$(C$_{2h}$)), **(c)** an example of E$_g$ phonon (at ~ 240 cm$^{-1}$) intensity experiencing a clear discontinuity in the LR channel but remaining absent in the LR channel, corresponding to a E$_g$(C$_{3i}$) phonon transforming into a B$_g$(C$_{2h}$) mode, and **(d)** M$_2$ mode intensity. Inset shows a zoom-in between 1.5 and 2.5 T with both increasing (+B) and decreasing (−B) magnetic fields. Error bars correspond to one standard error of the fitted peak intensities. **(e)** Schematic illustration for the shearing of CrI$_3$ layers across the magnetic phase transition $B_c$. The light blue ellipses represent the directors between layers formed because of this lattice deformation.

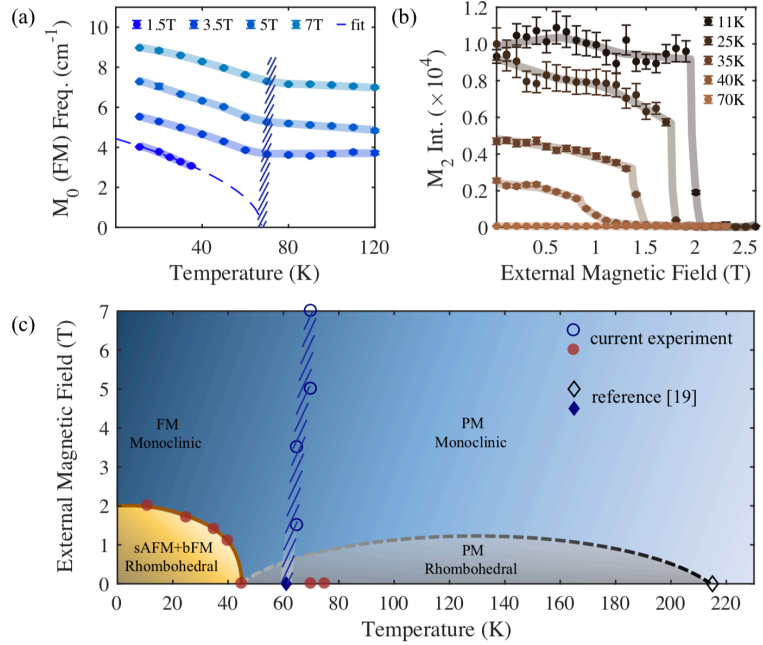

**Figure 4** Constructing the temperature v.s. external magnetic field phase diagram for 3D $CrI_3$. **(a)** Temperature dependence of the $M_{0c}$ frequency for the FM spin wave branch at multiple magnetic fields of 1.5 T (below $B_c$), 3.5 T, 5 T, and 7 T (above $B_c$). The dashed line is a fit to the data taken at 1.5 T while the solid lines are guide to the eyes. The striped pattern highlights the crossover between the paramagnetism (PM) and ferromagnetism (FM). **(b)** Magnetic field dependence of $M_2$ intensity at multiple temperatures of 11 K, 25 K, 35 K, 40 K (below $T_c$), and 70 K (above $T_c$). Error bars correspond to one standard error of the fitted parameters. **(c)** A temperature v.s. external magnetic phase diagram constructed based on both the findings from the current work (blue open and brown filled circles) and the literature knowledge (blue open and filled diamonds, Ref. [22]). The gray solid line represents a true phase boundary between the mixed magnetic phase with a rhombohedral structure and the FM phase with a monoclinic structure. The gray dashed line is for a potential phase boundary between the rhombohedral and monoclinic PM phases. The blue stripped pattern indicates the crossover between the PM and FM phases.

**References**

1    Gong, C. *et al.* Discovery of intrinsic ferromagnetism in two-dimensional van der Waals crystals. *Nature* **546**, 265, doi:10.1038/nature22060 (2017).

2    Huang, B. *et al.* Layer-dependent ferromagnetism in a van der Waals crystal down to the monolayer limit. *Nature* **546**, 270, doi:10.1038/nature22391 (2017).

3    Zhong, D. *et al.* Van der Waals engineering of ferromagnetic semiconductor heterostructures for spin and valleytronics. *Science Advances* **3**, e1603113, doi:10.1126/sciadv.1603113 (2017).

4    Deng, Y. *et al.* Gate-tunable room-temperature ferromagnetism in two-dimensional $Fe_3GeTe_2$. *Nature* **563**, 94-99, doi:10.1038/s41586-018-0626-9 (2018).

5    Huang, B. *et al.* Electrical control of 2D magnetism in bilayer $CrI_3$. *Nature Nanotechnology* **13**, 544-548, doi:10.1038/s41565-018-0121-3 (2018).

6    Jiang, S., Li, L., Wang, Z., Mak, K. F. & Shan, J. Controlling magnetism in 2D $CrI_3$ by electrostatic doping. *Nature Nanotechnology* **13**, 549-553, doi:10.1038/s41565-018-0135-x (2018).

7    Jiang, S., Shan, J. & Mak, K. F. Electric-field switching of two-dimensional van der Waals magnets. *Nature Materials* **17**, 406-410, doi:10.1038/s41563-018-0040-6 (2018).

8    Klein, D. R. *et al.* Probing magnetism in 2D van der Waals crystalline insulators via electron tunneling. *Science* **360**, 1218, doi:10.1126/science.aar3617 (2018).

9    Song, T. *et al.* Giant tunneling magnetoresistance in spin-filter van der Waals heterostructures. *Science* **360**, 1214, doi:10.1126/science.aar4851 (2018).

10   Huang, B. *et al.* Tuning inelastic light scattering via symmetry control in the two-dimensional magnet CrI3. *Nature Nanotechnology*, doi:10.1038/s41565-019-0598-4 (2020).

11   McCreary, A. *et al.* Distinct magneto-Raman signatures of spin-flip phase transitions in CrI3. *arXiv preprint arXiv:1910.01237* (2019).

12   Sun, Z. *et al.* Giant nonreciprocal second-harmonic generation from antiferromagnetic bilayer CrI3. *Nature*, doi:10.1038/s41586-019-1445-3 (2019).

13   Zhang, Y. *et al.* Magnetic Order-Induced Polarization Anomaly of Raman Scattering in 2D Magnet CrI3. *Nano Letters* **20**, 729-734, doi:10.1021/acs.nanolett.9b04634 (2020).

14   Li, T. *et al.* Pressure-controlled interlayer magnetism in atomically thin CrI3. *Nature Materials* **18**, 1303-1308, doi:10.1038/s41563-019-0506-1 (2019).

15   Song, T. *et al.* Switching 2D magnetic states via pressure tuning of layer stacking. *Nature Materials* **18**, 1298-1302, doi:10.1038/s41563-019-0505-2 (2019).

16   Niu, B. *et al.* Coexistence of Magnetic Orders in Two-Dimensional Magnet CrI3. *Nano Letters* **20**, 553-558, doi:10.1021/acs.nanolett.9b04282 (2020).

17  Kim, H. H. *et al.* One Million Percent Tunnel Magnetoresistance in a Magnetic van der Waals Heterostructure. *Nano Letters* **18**, 4885-4890, doi:10.1021/acs.nanolett.8b01552 (2018).

18  Wang, Z. *et al.* Very large tunneling magnetoresistance in layered magnetic semiconductor $CrI_3$. *Nature Communications* **9**, 2516, doi:10.1038/s41467-018-04953-8 (2018).

19  Lee, J.-U. *et al.* Ising-Type Magnetic Ordering in Atomically Thin FePS3. *Nano Letters* **16**, 7433-7438, doi:10.1021/acs.nanolett.6b03052 (2016).

20  Kim, K. *et al.* Antiferromagnetic ordering in van der Waals 2D magnetic material MnPS3 probed by Raman spectroscopy. *2D Materials* **6**, 041001, doi:10.1088/2053-1583/ab27d5 (2019).

21  Kim, K. *et al.* Suppression of magnetic ordering in XXZ-type antiferromagnetic monolayer NiPS3. *Nature Communications* **10**, 345, doi:10.1038/s41467-018-08284-6 (2019).

22  McGuire, M. A., Dixit, H., Cooper, V. R. & Sales, B. C. Coupling of Crystal Structure and Magnetism in the Layered, Ferromagnetic Insulator $CrI_3$. *Chemistry of Materials* **27**, 612-620, doi:10.1021/cm504242t (2015).

23  Chen, L. *et al.* Topological Spin Excitations in Honeycomb Ferromagnet $CrI_3$. *Physical Review X* **8**, 041028, doi:10.1103/PhysRevX.8.041028 (2018).

24  Jin, W. *et al.* Raman fingerprint of two terahertz spin wave branches in a two-dimensional honeycomb Ising ferromagnet. *Nature Communications* **9**, 5122, doi:10.1038/s41467-018-07547-6 (2018).

25  Seyler, K. L. *et al.* Ligand-field helical luminescence in a 2D ferromagnetic insulator. *Nature Physics* **14**, 277-281, doi:10.1038/s41567-017-0006-7 (2018).

26  Fleury, P. A. & Loudon, R. Scattering of Light by One- and Two-Magnon Excitations. *Physical Review* **166**, 514-530, doi:10.1103/PhysRev.166.514 (1968).

27  Hayes, W. & Loudon, R. *Scattering of Light by Crystals*. (Courier Corporation, 2012).

28  Ji, J. *et al.* Giant magneto-optical Raman effect in a layered transition metal compound. *Proceedings of the National Academy of Sciences* **113**, 2349, doi:10.1073/pnas.1601010113 (2016).

29  Shcherbakov, D. *et al.* Raman Spectroscopy, Photocatalytic Degradation, and Stabilization of Atomically Thin Chromium Tri-iodide. *Nano Letters* **18**, 4214-4219, doi:10.1021/acs.nanolett.8b01131 (2018).

30  Djurdjić-Mijin, S. *et al.* Lattice dynamics and phase transition in $CrI_3$ single crystals. *Physical Review B* **98**, 104307, doi:10.1103/PhysRevB.98.104307 (2018).

31  Webster, L., Liang, L. & Yan, J.-A. Distinct spin–lattice and spin–phonon interactions in monolayer magnetic $CrI_3$. *Physical Chemistry Chemical Physics* **20**, 23546-23555, doi:10.1039/C8CP03599G (2018).

32   Larson, D. T. & Kaxiras, E. Raman spectrum of CrI$_3$: An *ab initio* study. *Physical Review B* **98**, 085406, doi:10.1103/PhysRevB.98.085406 (2018).

33   Ashcroft, N. W. & Mermin, N. D. *Solid State Physics*.  (Holt, Rinehart and Winston, 1976).

34   Tolédano, J. C. & Tolédano, P. *The Landau Theory of Phase Transitions: Application to Structural, Incommensurate, Magnetic and Liquid Crystal Systems*.  (World Scientific Publishing Company, 1987).

35   Johnston, D. C. Magnetic dipole interactions in crystals. *Physical Review B* **93**, 014421, doi:10.1103/PhysRevB.93.014421 (2016).

36   Sivadas, N., Okamoto, S., Xu, X., Fennie, C. J. & Xiao, D. Stacking-Dependent Magnetism in Bilayer CrI$_3$. *Nano Letters* **18**, 7658-7664, doi:10.1021/acs.nanolett.8b03321 (2018).

37   Chaikin, P. M. & Lubensky, T. C. *Principles of Condensed Matter Physics*.  (Cambridge University Press, 2000).

38   Cardy, J., Goddard, P. & Yeomans, J. *Scaling and Renormalization in Statistical Physics*. (Cambridge University Press, 1996).

Supplemental Material for

# Magnetic field-induced quantum phase transitions in a van der Waals magnet

Siwen Li[1,+], Zhipeng Ye[2,+], Xiangpeng Luo[1,+], Gaihua Ye[2], Hyun Ho Kim[3], Bowen Yang[3], Shangjie Tian[4], Chenghe Li[4], Hechang Lei[4], Adam W. Tsen[3], Kai Sun[1], Rui He[2,*], and Liuyan Zhao[1,*]

[1]Department of Physics, University of Michigan, 450 Church Street, Ann Arbor, Michigan 48109, USA

[2]Department of Electrical and Computer Engineering, 910 Boston Avenue, Texas Tech University, Lubbock, Texas 79409, USA

[3]Institute for Quantum Computing, Department of Chemistry, and Department of Physics and Astronomy, University of Waterloo, Waterloo, 200 University Ave W, Ontario, N2L 3G1, Canada

[4]Department of Physics and Beijing Key Laboratory of Opto-electronic Functional Materials & Micro-nano Devices, Remin University of China, Beijing 100872, China

[+] authors contributed equally;

[*] corresponding to: rui.he@ttu.edu and lyzhao@umich.edu

**Table of Contents**

S1. Raman selection rules for $CrI_3$ with the circular polarization basis

S2. Extended data of magnetic field dependent Raman spectra on 3D $CrI_3$ bulk

S3. Magnetic field dependent spin waves in $CrI_3$ flakes with interlayer AFM

S4. Calculations of magnetic energy for individual $Cr^{3+}$ from magnetic dipole interactions

S5. Magnetization v.s. magnetic field measurements in 3D $CrI_3$ bulk

S6. Spin wave calculations for the sAFM and bFM mixed state and its magnetic field dependence

S7. Comparison between the magnetic field- and temperature-induced monoclinic phases

S8. Magnetic field dependence of $M_1$ intensity

S9. Raman scattering analysis for $M_1$ and $M_2$

S10. Magnetic field-induced structural phase transition in $CrI_3$ flakes with interlayer AFM

## S1. Raman selection rules for CrI$_3$ with the circular polarization basis

Under magnetic fields, the quartz-based optical components manifest the magneto-optical Faraday effect by rotating the electric field polarization of linearly polarized light, as is the case for a transmission objective in proximity to a strong magnet in magnetic field dependent micro-Raman measurements. Such a Faraday effect introduces artifacts in Raman symmetry analysis based on linearly polarized light. As circularly polarized light is free of the impact from such a Faraday effect, we have used the circularly polarized light to ensure the accuracy of symmetry analysis in this Raman experiment on CrI$_3$ at various magnetic fields.

Three-dimensional (3D) CrI$_3$ bulk crystal possesses either the rhombohedral (space group #148, $R\bar{3}$, point symmetry $C_{3i}$) or the monoclinic (space group #12, $C2/m$, point symmetry $C_{2h}$) structures in different regimes of the temperature v.s. magnetic field phase diagram (Fig. 4c in the main text). In our normal-incident and backscattering geometry, the sample surface normal, denoted as $\hat{z}$ direction, coincides with the propagation directions of both the incident and scattered light, as well as the $c$-axis of the sample in its $C_{3i}$ rhombohedral phase. The Raman tensors for $C_{3i}$ point group in such a coordinate system take the forms of (showing only the $x$-$y$ components)

$$\chi(A_g) = \begin{pmatrix} a & 0 \\ 0 & a \end{pmatrix}, \qquad \chi(E_g) = \begin{pmatrix} b & c \\ c & -b \end{pmatrix}. \tag{1}$$

The shearing between CrI$_3$ layers reduces the point symmetry group of CrI$_3$ to the monoclinic one $C_{2h}$ with the principal $C_2$ axis being in-plane. This yields the Raman tensors

$$\chi(A_g) = \begin{pmatrix} a & 0 \\ 0 & b \end{pmatrix}, \qquad \chi(B_g) = \begin{pmatrix} 0 & c \\ c & 0 \end{pmatrix}. \tag{2}$$

The Raman intensity is evaluated by $I \propto \left| E_s^\dagger \cdot \chi \cdot E_i \right|^2$, where $E_i$ and $E_s$ are the electric polarization vectors of the incident and scattered light, respectively. Circular polarization vectors for the normal incidence geometry are written as $\hat{L} = \begin{pmatrix} 1 \\ i \end{pmatrix}$, $\hat{R} = \begin{pmatrix} 1 \\ -i \end{pmatrix}$ in the Cartesian basis. Substituting the Raman tensors above we obtain the selection rules in Table. S1. Note that in the rhombohedral phase $A_g$ and $E_g$ modes are expected to appear exclusively in parallel and crossed channels, respectively, whereas in the monoclinic phase $A_g$ modes show up in both channels and the $B_g$ modes only in the crossed channel.

As elaborated in Section S8, the antiferromagnetism (AFM) related Raman modes $M_{1,2}$ in the rhombohedral phase are anti-symmetric, with their Raman tensors being of the form

$$\chi(M_{1,2}) = \begin{pmatrix} 0 & a \\ -a & 0 \end{pmatrix}. \tag{3}$$

Such modes are non-vanishing only in the LL channel.

|  | LL | LR |
| --- | --- | --- |
| Rhombohedral/$C_{3i}$ | $A_g$, $M_{1,2}$ | $E_g$ |
| Monoclinic/$C_{2h}$ | $A_g$ | $A_g$, $B_g$ |

**Table. S1.** Raman selection rules for $CrI_3$ in the rhombohedral ($C_{3i}$) and monoclinic ($C_{2h}$) structures in circularly parallel (LL) and crossed (LR) channels.

**S2.    Extended data of magnetic field dependent Raman spectra on 3D $CrI_3$ bulk**

S2.1    Magnetic field dependent Raman spectra of 3D $CrI_3$ bulk in the LR channel at 10 K.

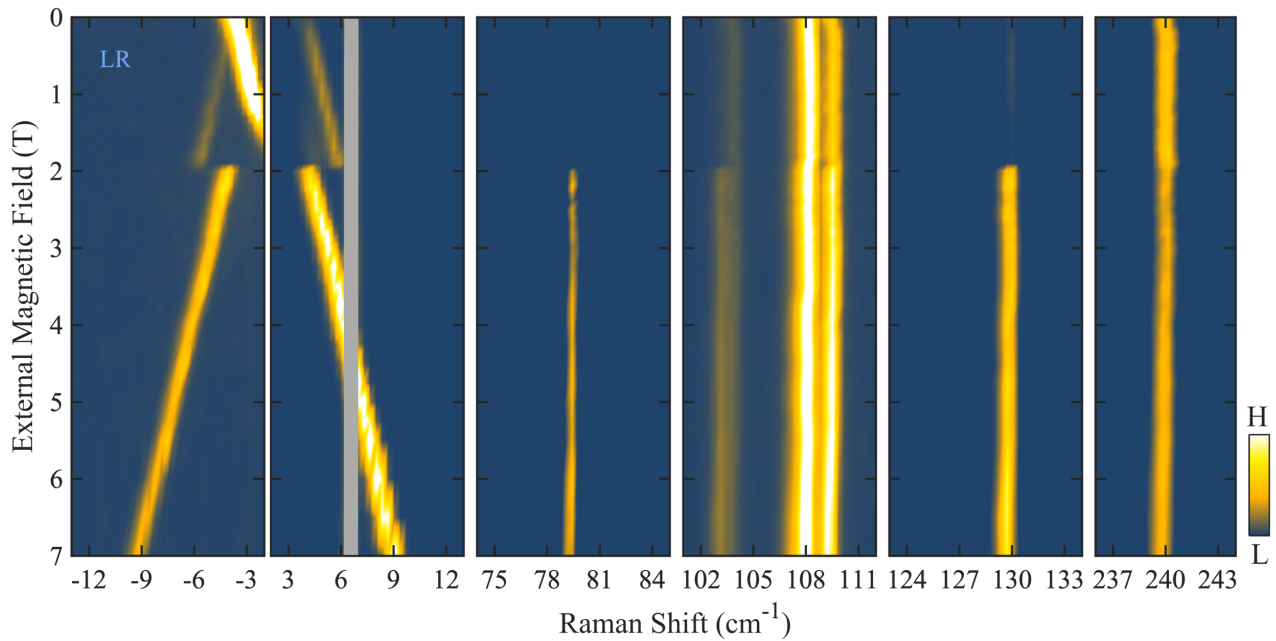

**Fig. S1    Magnetic field dependent Raman spectra of 3D CrI₃ bulk in the LR channel at 10 K**.

S2.2    Magnetic field dependent spin wave branches in 3D $CrI_3$ bulk in both Stokes and anti-Stokes sides and in both the LL and LR channels

The false color maps of magnetic field dependent Raman spectra in Fig. 1b in the main text and Fig. S1 above were plotted to have individual panels normalized to their own maxima, in order to make weak features more visible within each panel. However, we cannot make a direct comparison of Raman intensities across different panels. In Fig. S2 below, we provide color maps of magnetic field dependent spin wave modes in Fig. 1b and Fig. S1 using the same color scale so as to make a fair comparison between the spin wave branch intensities of Stokes and anti-Stokes sides in the LL and LR channels.

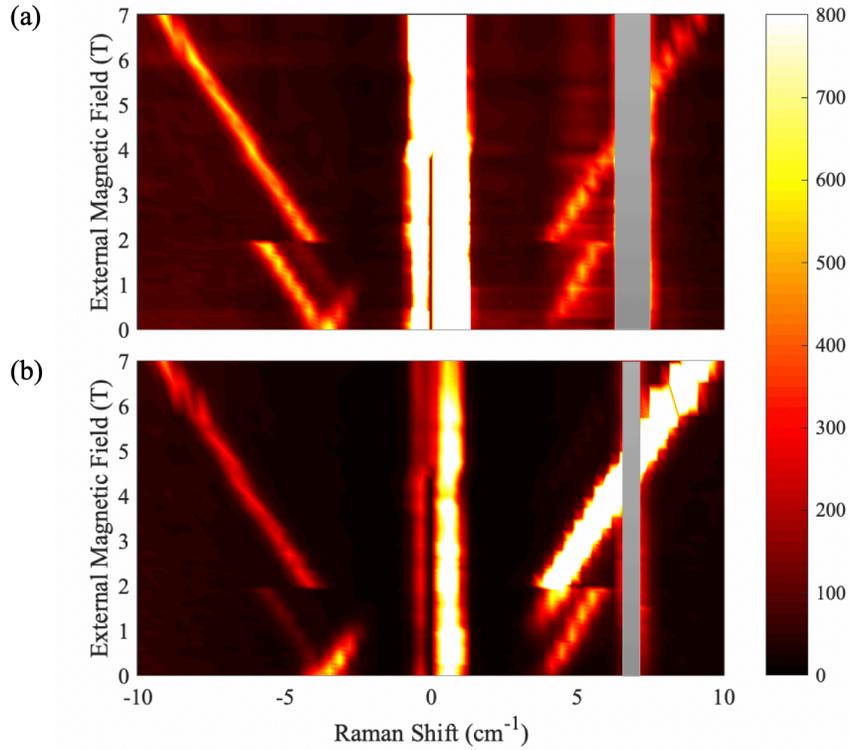

**Fig. S2  Color maps for the magnetic field-dependent Raman spectra.** The data were acquired in (**a**) the LL, and (**b**) the LR channel at 11 K. The color scale is set the same for the anti-Stokes and Stokes sides for both the LL and LR channels.

S2.3  Magnetic field dependent resonant Raman spectra on three freshly cleaved 3D $CrI_3$ bulk

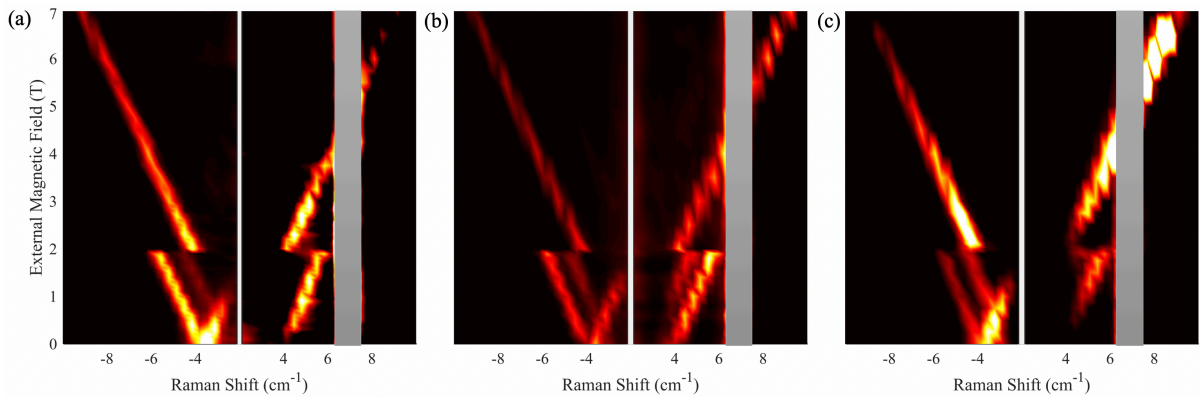

**Fig. S3  Magnetic field dependent resonant Raman spectra on three freshly cleaved 3D $CrI_3$ bulk.** (**a**) on 2019/04/23, (**b**) on 2019/10/10, and (**c**) on 2018/12/07. All three data sets show consistent experimental results of a mixture of sAFM and bFM in 3D $CrI_3$ bulk.

S2.4  Comparison of the magnetic field dependent Raman spectra between the increasing and decreasing magnetic field.

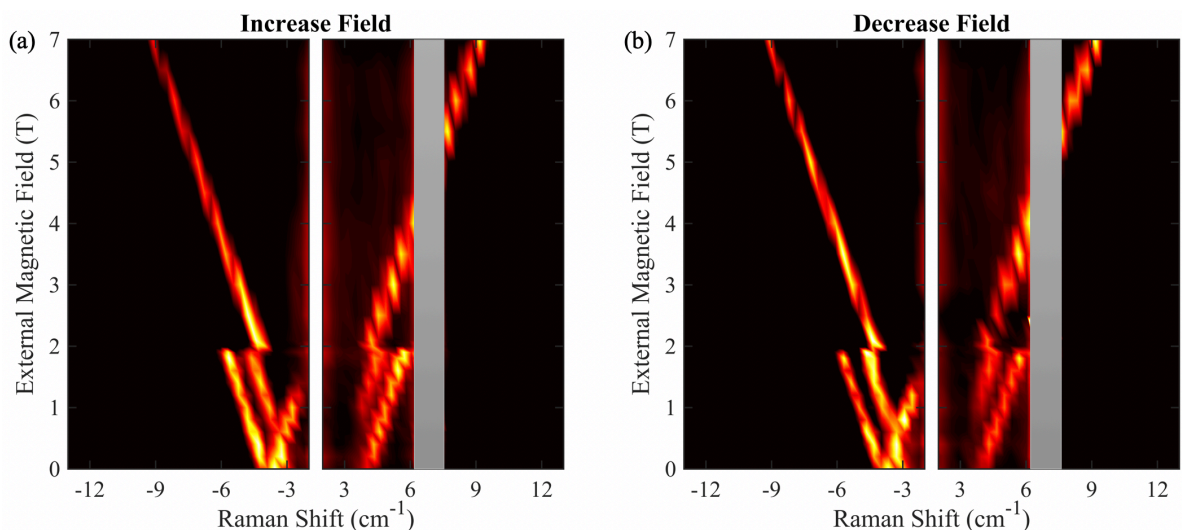

**Fig. S4  Comparison of the magnetic field dependent Raman spectra between increasing and decreasing magnetic field.** Data taken with (**a**) the increasing magnetic field, and (**b**) the decreasing magnetic field.

## S3.  Magnetic field dependent spin waves in $CrI_3$ flakes with interlayer AFM

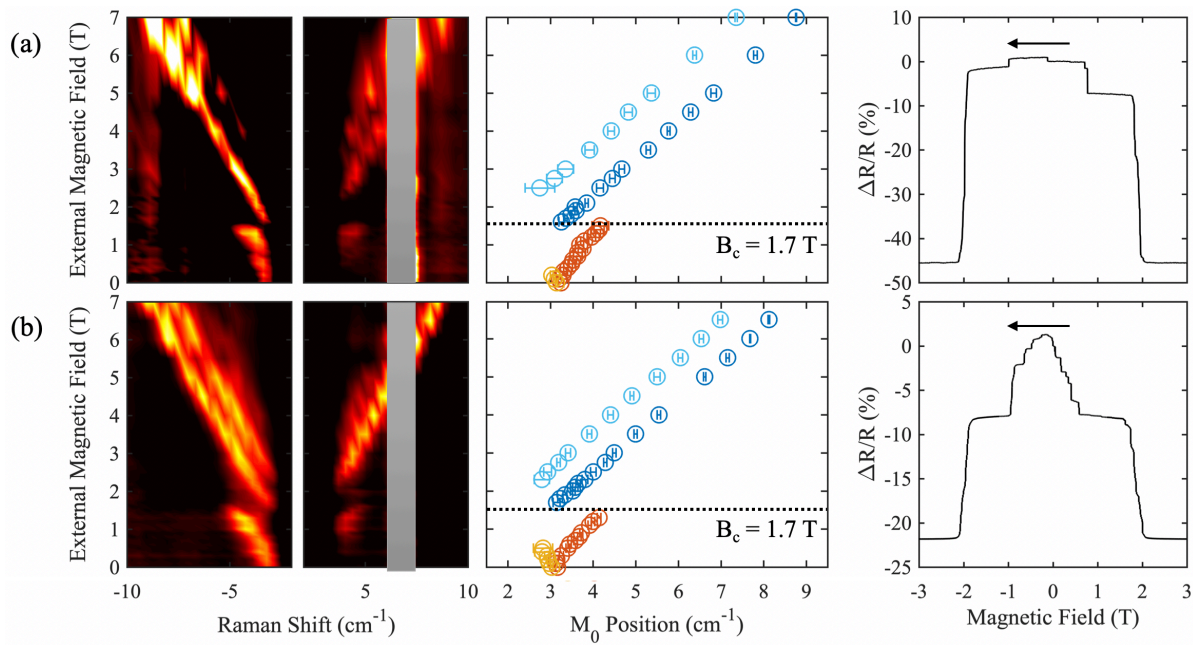

**Fig. S5  Resonant Raman spectroscopy data on two $CrI_3$ flakes with different thicknesses, together with their magneto-tunneling resistivity spectra.** Raw magnetic field-dependent magnon

Raman spectra maps, the plots of the fitted magnon frequencies v.s. magnetic fields, and magento-tunneling resistivity spectra for (**a**) 8-layer-thick $CrI_3$ flake, (**b**) 20-layer-thick $CrI_3$ flake. Below $B_c$, we can see the interlayer AFM spin wave branches (yellow and orange) in both samples, but the interlayer FM spin wave branch is missing in both samples. Above $B_c$, there are two FM spin wave branches in both flake samples, in contrast to the single FM spin wave branch in the 3D $CrI_3$ bulk. This is because the thin flakes break translational symmetry along the *c*-axis (the out-of-plane direction) and therefore result in satellite modes near the primary one, which can be seen for phonon modes as well and have been discussed in *Nature Communications* **9**, 5122 (2018). The magneto-tunneling resistivity measurements for both thin layer samples show a critical magnetic field of $B_c \sim 1.7$ T, which is consistent with that found in resonant Raman spectroscopy.

### S4. Calculations of magnetic energy for individual $Cr^{3+}$ from magnetic dipole interactions

Why do we have such spin mixtures in $CrI_3$? We distill this question to be Option1: "whether the surface AFM and bulk FM are both dominated by short-range exchange coupling" or Option 2: "whether the surface AFM is from the short-range exchange coupling while the bulk FM results from the long-range dipole-dipole interactions". Option 1 naturally requires different interlayer exchange coupling at the surface and in the bulk, which necessarily needs a surface structural reconstruction to realize. Option 2 explores the possibility of magnetic energetics, in particular under the assumption that surface and bulk share the same crystal structure and thus have identical exchange coupling and other coupling coefficients. In this option, we take into account the long-range magnetic dipole-dipole interactions, which favors the FM order and competes with the short-range interlayer AFM exchange coupling. In systems where this dipole interaction strength accumulates over a certain thickness to overcome the interlayer AFM coupling, magnetic dipole interaction dominated FM ground states can be realized in the bulk[1].

To test whether Option 2 is plausible in 3D $CrI_3$, we therefore performed a calculation on the long-range magnetic dipole-dipole interaction for the $CrI_3$ with the rhombohedral crystal structure both at the surface and in the bulk with the lattice constant adopted from literature. We calculated the magnetic energy of a single magnetic moment at the surface interacting with all the rest of magnetic dipole moments within a cylinder (whose radius $R$ is 3 $\mu$m, a typical in-plane ferromagnetic domain size, and depth is labeled by the number of layers $N$) through dipole-dipole interactions.

$$E_{d-d} = \sum_{\substack{i,j,k \\ (r_{ijk})_\perp < R}} -\frac{\mu_0}{4\pi |\vec{r}_{ijk}|^3} \left(3(\vec{m}_{ijk} \cdot \hat{r}_{ijk})(\vec{m}_{000} \cdot \hat{r}_{ijk}) - \vec{m}_{ijk} \cdot \vec{m}_{000}\right)$$

where $i, j, k$ are the lattice indices for the $Cr^{3+}$ ion sites, the single $Cr^{3+}$ site of interest is labeled as (0,0,0) because it's in the center of the top surface of the cylinder, $\vec{r}_{ijk}$ is the separation vector from the central $Cr^{3+}$ at (0,0,0) to the $Cr^{3+}$ at $(i,j,k)$, and $(r_{ijk})_\perp$ is the projected length on to the sample surface plane.

There is no simple analytical solution to this dipole-dipole interaction, and therefore, we performed computational calculations. The result is plotted in Fig. R1 below. We can see that the magnetic energy per single $Cr^{3+}$ ion from long-range dipole-dipole interactions is only $< 1\ \mu eV$ for 20 layers of $CrI_3$ (20 layers are the estimated thickness of the surface AFM phase). This magnetic energy scale is over two orders of magnitude weaker than the interlayer AFM exchange couple of $150\ \mu eV$. Therefore, the long-range magnetic dipole interaction is not sufficient to overcome the short-range interlayer AFM exchange coupling so as to achieve the interlayer FM phase in regions of 3D $CrI_3$ about 20 layers from the surface. This finding is consistent with the literature on dipole-dipole interaction dominated magnets (whose leading energy scale is typically in order of $\mu eV$ and transition temperature is sub-K)

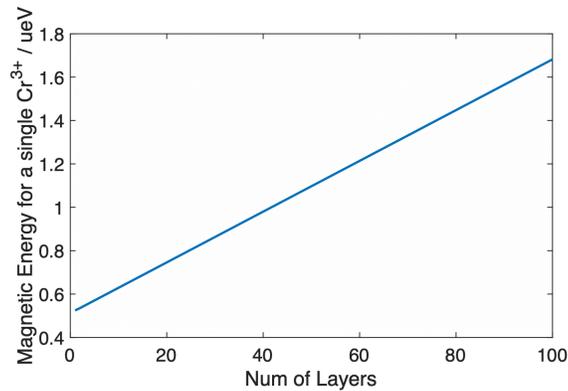

**Fig. R1** Layer number dependence of the magnetic energy for a single $Cr^{3+}$ magnetic moment, through long-range dipole-dipole interactions.

Our calculations rule out Option 2, and therefore suggest Option 1 (surface reconstruction) as a likely reason for achieving surface AFM and bulk FM.

### S5.  Magnetization v.s. magnetic field measurements in 3D $CrI_3$ bulk

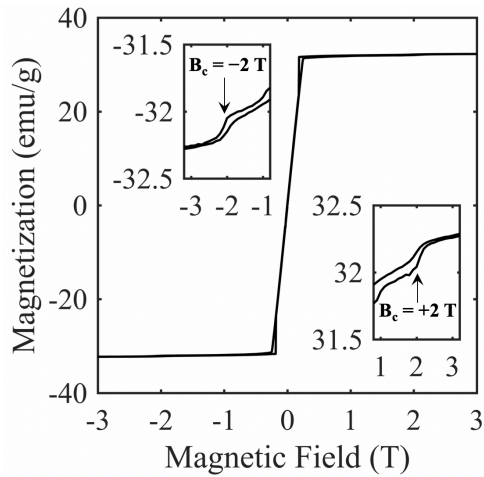

**Fig. S7    The plot of the magnetization v.s. out-of-plane magnetic field data taken on 3D CrI₃ bulk.** The main panel shows the clear signature of the poling field of the bFM domain states, $B_p$ ~ 0.1 T. The two insets show the close-look around $B_c$ = ± 2T, the critical field for the sAFM to FM transition, in which a weak but observable anomaly in the magnetization is seen.

With these additional results from S3 and S4 sections above, we provide an in-depth analysis that confirms the mixed state of sAFM and bFM in 3D CrI₃.

*First, we establish the interlayer AFM state in addition to the FM state in 3D CrI₃.*

We summarize the experimental findings point-by-point as follows:

1. Through the magneto-tunneling resistance measurements, we find that the CrI₃ flakes of the thickness used in our experiment have interlayer AFM ordering, as evidenced by the extremely low magneto-tunneling resistance at low magnetic fields. Across the critical field $B_c$, about 1.7 T, the interlayer AFM transits to the FM phase, which is supported by the significant reduction of magneto-tunneling resistance above $B_c$.

2. Through the magnetization measurements, we find that the bulk crystal is primarily in the FM order, and the saturation poling field $B_p$ is as low as ~ 0.1 T. This is to say that the magnetic field required to polarize the FM domains is as low as ~ 0.1 T. In addition, there is a very small but observable signature of the interlayer AFM to FM transition at ~ ± 2 T (Fig. S7).

3. Through the magnetic field dependent Raman spectroscopy measurements of CrI₃ flakes, both samples that are interlayer AFMs show **a)** a magnetic phase transition at a critical field $B_c$ ~ 1.7 T, which is consistent with the interlayer AFM to FM transition detected by the magneto-tunneling resistance measurements; **b)** two spin wave branches with opposite magnetic field dependence below $B_c$, which corroborates with the interlayer AFM phase; **c)** the absence of the third spin wave branch below $B_c$ that is present in 3D CrI₃, which is again consistent with the surface interlayer AFM phase.

Comparing the three experiments above in both bulk and multilayer CrI₃, we can confidently conclude: We find in bulk CrI₃, the FM phase for a great majority of the bulk volume, clear signatures of the interlayer AFM phase, and its phase transition into the FM phase at $B_c$. These layered AFM-related behaviors are exactly the same as those in CrI₃ flakes known to be interlayer AFMs. At the same time, we also exclude the possibilities of spin waves from FM domains with opposite magnetization as it only takes ~ 0.1 T to polarize the FM domains whereas we see the three spin wave branches in 3D CrI₃ surviving up to $B_c$ ~ 2 T.

*Second, we answer the question where the interlayer AFM happens in 3D CrI₃ bulk.*

Resonant Raman spectroscopy measurements are subject to finite penetration depth, and therefore only survey a finite number of layers from the top surface. We have measured more than ten bulk CrI₃ samples with *freshly cleaved surfaces* and have observed the very same signatures of interlayer AFM and magnetic transitions in every individual sample. Fig. S5 shows the results from two flake examples. Such a consistent observation, together with the very small volume of interlayer AFM in 3D CrI$_3$ bulk, suggests that the interlayer AFM has to happen at the top layers, rather than deep in the bulk.

**S6.    Spin wave calculations for the sAFM and bFM mixed state and its magnetic field dependence**

We have performed standard spin wave calculations for a 3D magnet made of ABC stacked 2D honeycomb magnetic lattices.

To calculate the spin wave dispersion for the sAFM state, we simulate with an AFM interlayer exchange coupling and FM intralayer exchange couplings, and the spin Hamiltonian can be written as

$$H = H_0 + H_I + H_f$$
$$H_0 = -\frac{1}{2}\sum_{l,<i,j>}[J_z S_{l,i}^z S_{l,j}^z + J_x(S_{l,i}^x S_{l,j}^x + S_{l,i}^y S_{l,j}^y)]$$
$$H_I = \frac{1}{2}J_I \sum_{i,<l,l'>} \vec{S}_{l,i} \cdot \vec{S}_{l',j} \quad (4)$$
$$H_f = -g\mu_B B \sum_{l,i} S_{l,i}^z$$

where $J_{x,z} > 0$ and $J_I > 0$ are intra- and inter- layer exchange couplings, respectively, $S_{l,i}^{x(y,z)}$ is the $x$ ($y$, $z$)-component of spin at site $i$ of layer $l$, $B$ is the external magnetic field aligned along the $z$-direction, and $H_0$, $H_I$, and $H_f$ are spin Hamiltonians for the intralayer FM interactions, intralayer AFM interactions, and Zeeman effect in CrI$_3$.

There are 4 different spin sites per magnetic unit cell in sAFM because of two sublattice per honeycomb lattice and two layers of honeycomb lattices per AFM unit cell. After applying the Holstein-Primakoff transformation and the Bogoliubov transformation, we found the magnon energies at the Γ-point are

$$E_1 = \frac{1}{2}S\sqrt{(J_z - J_x)z[4J_I + (J_z - J_x)z]} \pm g\mu_B B;$$
$$E_2 = \frac{1}{2}S\sqrt{(J_z + J_x)z[4J_I + (J_z + J_x)z]} \pm g\mu_B B. \quad (5)$$

Without the external magnetic field (*i.e.*, $B$ = 0 T), each branch is two-fold degenerated with one magnon carrying spin 1 and the other spin −1. The presence of an external field lifts this degeneracy, and therefore the two magnons of spin ±1 per branch undergo the Zeeman shifts in the opposite treands at the same rate (*i.e.*, slope of $\pm g\mu_B$).

In the mixed sAFM and bFM state of 3D CrI$_3$, the bFM provides an effective magnetic field $B_0$ to the sAFM even when there is no external magnetic field. This results in the slight splitting of the two magnons per branch at 0 T. In our experimental data, the magnon branch we observed is the acoustic branch $E_1$ whose field dependence is shown in Fig. 2a and labeled with the sAFM.

In the case of bFM below 2 T or FM above 2 T, we simulate the effective interlayer coupling with a FM one, and there are only 2 different spin sites because of two sublattices per honeycomb layer and one layer per FM unit cell. They correspond to one acoustic and one optical branch at 0 T, but are both spin -1 without any degeneracy. Experimentally, we observed the acoustic branch labeled as bFM in Fig. 2a. The interlayer exchange coupling $J_I$ is no longer the same as that in Eq. 2, but this has no impact on the magnetic field dependence of the acoustic branch which takes the following form of

$$E_1' = \frac{1}{2}Sz(J_z - J_x) + g\mu_B B. \qquad (6)$$

Comparing $E_1'$ and $E_1$, the magnon with spin -1 of bFM has smaller energy than that of sAFM that is indeed consistent with our experimental observation.

By comparing our calculated and experimental magnon energies, we have extracted the following important parameters

- Below 2 T:
  Interlayer exchange coupling: $J_I$ = 0.15 meV
  Intralayer exchange anisotropy: $J_z - J_x$ = 0.13 meV
  Effective magnetic field from the bFM at 0 T external field: $B_0$ = 0.32 T
- Above 2 T:
  Intralayer exchange anisotropy $J_z - J_x$ = 0.11 meV.

Given the fact that only the acoustic magnon branch is probed experimentally, we can only fit out the intralayer exchange anisotropy $J_z - J_x$, but not the individual values of $J_x$ and $J_z$.

It is worth mentioning that, apart from studying magnetic excitations, we can also estimate the interlayer exchange coupling $J_I$ using a ground state property, *i.e.* the magnetic phase transition field of 2 T. The AFM interlayer exchange coupling favors spins in adjacent layers aligning along opposite directions, whereas the external field favors all the spins aligning along the field direction. Using Eq. 2, the ground state energies of sAFM and FM per spin are

- sAFM:

$$H_I = -\frac{1}{2}S^2 J_I$$

$$H_f = 0$$

- FM:

$$H_I = \frac{1}{2}S^2 J_I$$

$$H_f = -g\mu_B B S$$

The phase transition happens when the total energy of FM becomes lower than that of the sAFM, that is $g\mu_B B_c S = S^2 J_I$. With $B_c = 2$ T, we get $J_I = 0.15$ meV, which corroborates the previous result of $J_I = 0.15$ meV obtained using magnon energies.

### S7. Comparison between the magnetic field- and temperature-induced monoclinic structure

We compare the Raman spectra between two cases, the magnetic field-induced monoclinic phase below $T_N$ above $B_c$ and the temperature-induced monoclinic phase above $T_s$ and at $B = 0$ T, as shown in Fig. S8 below. For the magnetic field-induced monoclinic phase, we can clearly observe the relaxation of selection rules from the rhombohedral point group, as the leakage of $A_g$ and $E_g$ modes appear in the LR and LL channel, respectively (Fig. S8(a)). In contrast, for the temperature-induced monoclinic phase, we barely find the selection rules relaxation from the rhombohedral $C_{3i}$ to the monoclinic $C_{2h}$ point group, because the $A_g$ and $E_g$ modes remain nearly absent in the LR and LL channel, respectively (Fig. S8(b)). It could be the weaker interlayer van der Waals coupling between layers in the temperature-induced monoclinic phase than in the magnetic field-induced monoclinic structure, that makes the in-plane phonon modes (*e.g.*, the $A_g$ and $E_g$ in the main text figure Fig. 3) hardly feel the symmetry reduction from the shearing between layers.

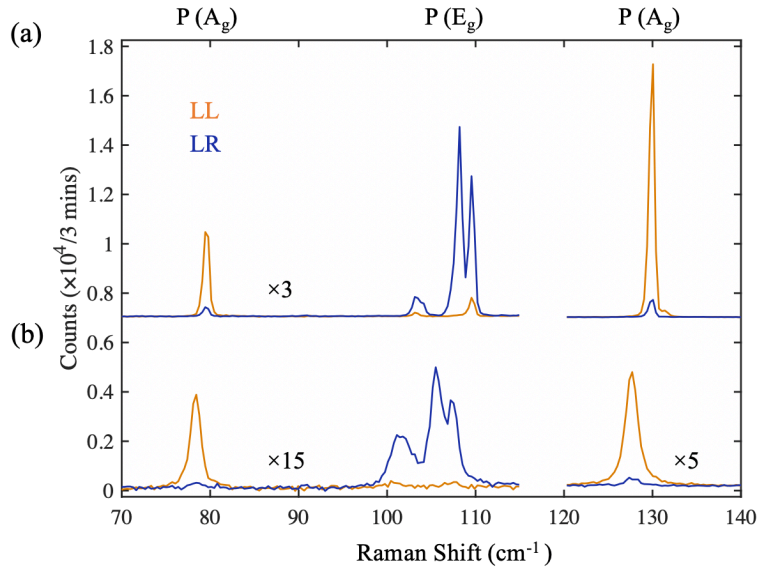

**Fig. S8  Comparison between the Raman spectra taken on the magnetic field- and temperature-induced monoclinic structure.** Raman spectra in both LL and LR channels taken on (**a**) magnetic field-induced monoclinic phase, $T$ = 11 K and $B$ = 3 T, where a clear relaxation of the selection rule from rhombohedral structure is observed (the leakage of $A_g$ from LL into LR channel, and the leakage of $E_g$ from LR to LL channel), and (**b**) temperature-induced monoclinic phase, $T$ = 290 K and $B$ = 0 T, where the selection rule relaxation is hardly observable.

## S8.  Magnetic field dependence of $M_1$ intensity

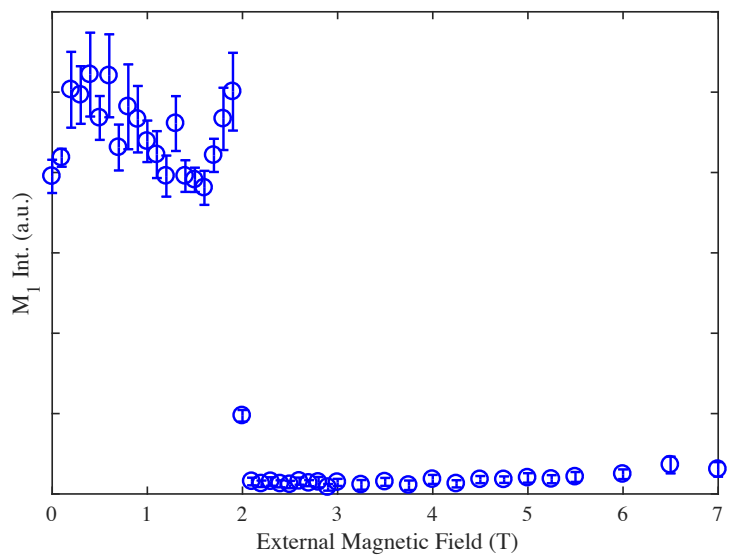

**Fig. S9**  Intensity of $M_1$ in LL channel as a function of external field at 11 K

## S9.  Raman scattering analysis for $M_1$ and $M_2$

One proposed origin of $M_1$ and $M_2$ is the paired excitation of each corresponding zone boundary phonon of the non-magnetic structure at $\vec{q} = (0, 0, k_c)$, $\Delta X_k^{\vec{q}}$, and the AFM order of $-\vec{q}$, $M_l^{-\vec{q}}$, which actually preserves the total momentum of zero so that it is detectable by Raman techniques. Please note that the relevant zone boundary phonons for $M_1$ and $M_2$ are different. This excitation can also be described by zone-folding by a single copy (odd numbers of copies) of magnetic order. We can further understand the selection rule of this collective excitation ($\Delta X_k^{\vec{q}} M_l^{-\vec{q}}$) by performing an expansion of the linear optical susceptibility ($\chi_{ij}$) tensor with respect to both the phonon vibration and the magnetization and find the induced change in linear optical susceptibility ($\Delta \chi_{ij}$) to be

$$\Delta\chi_{ij} = \left.\frac{\partial^2\chi_{ij}}{\partial X_k^{\vec{q}} \partial M_l^{-\vec{q}}}\right|_{\Delta X_k^{\vec{q}}=0, M_l^{-\vec{q}}=0} \Delta X_k^{\vec{q}} M_l^{-\vec{q}} = K_{ijkl}\Delta X_k^{\vec{q}} M_l^{-\vec{q}}. \qquad (7)$$

$K_{ijkl}$ is polar tensor that is invariant under the symmetry operations of the lattice point group of $C_{3i}$, with 27 independent tensor elements[2],

$$K_{ijkl} =$$

$$\left(\begin{pmatrix}\begin{pmatrix} K_{xxxx} & K_{xxxy} & K_{xxxz} \\ -K_{yyxy} & K_{xxyy} & -K_{yyyz} \\ K_{xxzx} & -K_{yyzy} & K_{yyzz} \end{pmatrix} \begin{pmatrix} -K_{yxyy} & K_{xyxy} & -K_{yyyz} \\ K_{xyyx} & K_{xyyy} & -K_{xxxz} \\ -K_{yyzy} & -K_{xxzx} & K_{xyzz} \end{pmatrix} \begin{pmatrix} K_{xzxx} & -K_{yzyy} & K_{xzxz} \\ -K_{yzyy} & -K_{xzxx} & K_{xzyz} \\ K_{yzzy} & K_{xzzy} & 0 \end{pmatrix} \\ \begin{pmatrix} -K_{xyyy} & K_{xyyx} & -K_{yyyz} \\ K_{xyxy} & K_{yxyy} & -K_{xxxz} \\ -K_{yyzy} & -K_{xxzx} & -K_{xyzz} \end{pmatrix} \begin{pmatrix} K_{xxyy} & K_{yyxy} & -K_{xxxz} \\ -K_{xxxy} & K_{xxxx} & K_{yyyz} \\ -K_{xxzx} & K_{yyzy} & K_{yyzz} \end{pmatrix} \begin{pmatrix} -K_{yzyy} & -K_{xzxx} & -K_{xzyz} \\ -K_{xzxx} & K_{yzyy} & K_{yzyz} \\ -K_{xzzy} & K_{yzzy} & 0 \end{pmatrix} \\ \begin{pmatrix} K_{zxxx} & -K_{zyyy} & K_{zyyz} \\ -K_{zyyy} & -K_{zxxx} & K_{zxyz} \\ K_{zyzy} & K_{zxzy} & 0 \end{pmatrix} \begin{pmatrix} -K_{zyyy} & -K_{zxxx} & -K_{zxyz} \\ K_{zxxx} & K_{zyyy} & K_{zyyz} \\ -K_{zxzy} & K_{zyzy} & 0 \end{pmatrix} \begin{pmatrix} K_{zzyy} & K_{zzxy} & 0 \\ -K_{zzxy} & K_{zzyy} & 0 \\ 0 & 0 & K_{zzzz} \end{pmatrix}\end{pmatrix}\right). \quad (8)$$

The Onsager relationship of $\Delta\chi_{ij}(\Delta\vec{X},\vec{M}) = \Delta\chi_{ji}(\Delta\vec{X},-\vec{M})$[3,4] further constraints the form of Eq. 8 and $\Delta\chi_{ij}(\Delta\vec{X},\vec{M})$ takes the following form

$$\Delta\chi_{xx} = \Delta\chi_{yy} = \Delta\chi_{zz} = 0;$$

$$\Delta\chi_{yx} = -\Delta\chi_{xy} = \frac{1}{2}[(X_yM_x - X_xM_y)(K_{xyxy} - K_{xyyx}) + (X_xM_x - X_yM_y)(K_{yxyy} - K_{xyyy})] - X_zM_zK_{xyzz};$$

$$\Delta\chi_{zx} = -\Delta\chi_{xz} = \frac{1}{2}[(X_xM_x - X_yM_y)(K_{zxxx} - K_{xzxx}) + (X_yM_x + X_xM_y)(K_{yzyy} - K_{zyyy}) + \cdots$$
$$X_xM_z(K_{zyyz} - K_{xzxz}) + X_yM_z(K_{zxyz} - K_{xzyz}) + X_zM_x(K_{zyzy} - K_{yzzy}) + X_zM_y(K_{zxzy} - K_{xzzy})]; \qquad (9)$$

$$\Delta\chi_{zy} = -\Delta\chi_{yz} = \frac{1}{2}[(X_xM_x - X_yM_y)(K_{yzyy} - K_{zyyy}) + (X_yM_x + X_xM_y)(K_{xzxx} - K_{zxxx}) + \cdots$$
$$X_xM_z(K_{xzyz} - K_{zxyz}) + X_yM_z(K_{zyyz} - K_{yzyz}) + X_zM_x(K_{xzzy} - K_{zxzy}) + X_zM_y(K_{zyzy} - K_{yzzy})].$$

As shown in Eq. 9, $\Delta\chi_{ij}$ is antisymmetric, so that the mode for such a composite object only shows up in the crossed channel with the linear polarization basis or the LL channel with the circular polarization basis.

**S10. Magnetic field-induced structural phase transition in CrI₃ flakes with interlayer AFM**

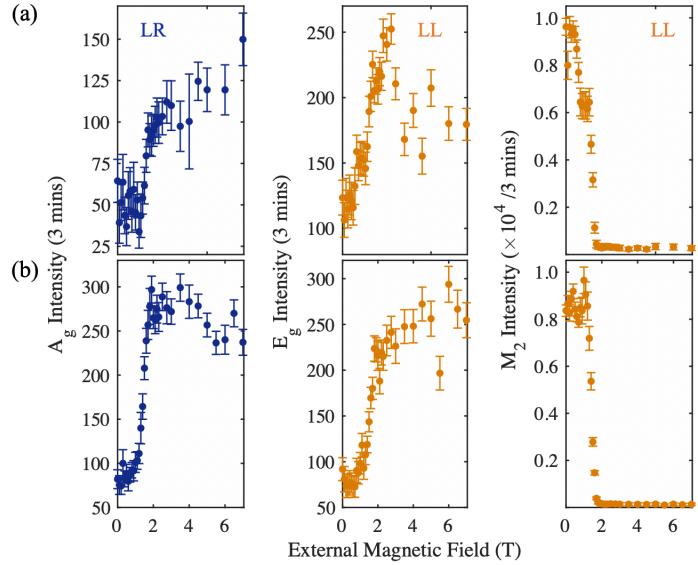

**Fig. S10 Selection rule relaxation for the phonon modes in CrI$_3$ flakes.** $A_g$ and $E_g$ mode intensity leakage to the opposite polarization channels and $M_2$ mode intensity as a function of the applied magnetic field in (**a**) 8-layer-thick, (**b**) 20-layer-thick CrI$_3$ flakes.

## References

1   Johnston, D. C. Magnetic dipole interactions in crystals. *Physical Review B* **93**, 014421, doi:10.1103/PhysRevB.93.014421 (2016).

2   Boyd, R. W. *Nonlinear optics*. (Elsevier, 2003).

3   Landau, L. D. *et al. Electrodynamics of continuous media*. Vol. 8 (elsevier, 2013).

4   Wettling, W., Cottam, M. & Sandercock, J. The relation between one-magnon light scattering and the complex magneto-optic effects in YIG. *Journal of Physics C: Solid State Physics* **8**, 211 (1975).